\documentclass[10pt]{article}

\usepackage{
graphicx,amsmath,amssymb,bm}

\newtheorem{prop}{Prop}[section]
\newtheorem{lemma}[prop]{Lemma}

\newtheorem{theorem}[prop]{Theorem}

\newtheorem{note}[prop]{Note}

\title{No  replica symmetry breaking  phase \\
in random field Ginzburg-Landau model 
 }

\author{C. Itoi \ and Y. Utsunomiya  \\
Department of Physics, GS and CST, Nihon University, \\
Kanda-Surugadai, Chiyoda, Tokyo 101-8308, Japan} 
\begin{document}
\maketitle
\begin{abstract}{
It is proved that the variance of spin overlap vanishes  in the infinite volume limit of 
 random field Ginzburg-Landau  model  
 using its FKG property. 
}
\end{abstract}


\section{Introduction}
Replica symmetry breaking phenomena in disordered spin systems have been studied extensively, 
since Talagrand proved Parisi's replica symmetry breaking formula \cite{Pr} for the Sherrington-Kirkpatrick model \cite{SK} in a mathematically rigorous manner \cite{T2}.
It is well known that the replica symmetry breaking appears generally in mean field disordered  spin models at low temperature
as a spontaneous symmetry breaking phenomenon.
 There have been lots of discussions whether or not, the random field Ising model  has some nontrivial phases due to its randomness such as
 replica symmetry breaking phase or spin glass phase.
 Krzakala,  Ricci-Tersenghi, Sherrington and Zdeborova have
 pointed out an evidence that spin glass phase does not exists in 
  the random field Ising model, the random field Ginzburg-Lamdau model and  random temperature Ginzburg-Landau model
  \cite{K,K2}, which satisfy  the  Fortuin-Kasteleyn-Ginibre (FKG) inequality \cite{FKG}. 
Recently, Chatterjee has proved  the absence of replica symmetry breaking in the random field Ising model in an arbitrary dimension rigorously  \cite{C2}.
He  proved that the variance of the spin overlap vanishes almost everywhere in the coupling constant space of the random field Ising model
using  the FKG inequality \cite{FKG} and the Ghirlanda-Guerra identities.  His method has been extended to quantum systems with the weak FKG property \cite{I3}. 
In  the present paper, we prove that the replica symmetry breaking does not occur also in  the random field  Ginzburg-Landau model, as well as the random field Ising model.\\

First, we define the model and several functions. 
Coupling constants  in a system with quenched disorder are given by  i.i.d. random variables.
We can regard a given disordered sample as a system obtained by a random sampling of these variables.
All physical quantities in such systems are functions of these random variables.
Consider a disordered Ginzburg-Landau model on a $d$ dimensional 
 hyper cubic lattice  $\Lambda_L:= [1,L]^d \cap {\mathbb Z}^d$ whose volume is $|\Lambda_L|=L^d$.
Let  $J=(J_{x,y})_{x,y\in \Lambda}$ be a real symmetric matrix such that $J_{x,y}=1$, if $|x-y|=1$, otherwise $J_{x,y}=0$. 
Define  Hamiltonian as a function of spin configuration $\phi=(\phi_x)_{x \in \Lambda_L} \in {\mathbb R}^{\Lambda_L}$
by
$$
H(\phi,g) :=-\sum_{|x,y \in \Lambda_L} J_{x,y}  \phi_x \phi_y  -h \sum_{x\in \Lambda_L}g_x \phi_x,
$$ 
where a vector  $g=(g_x)_{ x \in \Lambda_L}$  consists of
 i.i.d. standard Gaussian random variables  with a real constant $ h$.
Here, we define Gibbs state for the Hamiltonian.
For a positive $\beta $,  the  partition function is defined by
\begin{equation}
Z_L(\beta, u,h,r,g) :=  \int_{{\mathbb R}^{|\Lambda_L|}} D \phi e^{ - \beta H(\phi,g)},
\end{equation}
where the measure $D\phi$ is defined by
$$
D \phi = \prod_{x \in \Lambda_L} d \phi_x e^{-u \phi_x^4 + r \phi_x^2}
$$
for $u > 0$ and $r \in {\mathbb R}.$
The  expectation of a function of spin configuration $f(\phi)$ in the Gibbs state is given by
\begin{equation}
\langle f(\phi) \rangle =\frac{1}{Z_L(\beta,h,u,r,g)} \int_{{\mathbb R}^{|\Lambda_L|}} D\phi f(\phi)  e^{ - \beta H(\phi,g)}.
\end{equation}

Define the following functions of  $(\beta, h,u,r) \in [0,\infty)^3 \times {\mathbb R}$ and randomness
$g=(g_x)_{x \in \Lambda_L}$
\begin{equation}
\psi_L(\beta, h,u,r,g) := \frac{1}{|\Lambda_L|} \log Z_L(\beta,h,u,r,g), \\ 
\end{equation}
$-\frac{|\Lambda_L|}{\beta}\psi_L(\beta,h,u,r,g)$ is called free energy in statistical physics.
We define a function $p_L:[0,\infty)^3 \times {\mathbb R} \rightarrow {\mathbb R}$ by
\begin{eqnarray}
p_L(\beta,h,u,r):={\mathbb E} \psi_L(\beta,h,u,r,g) ,
\end{eqnarray}
where ${\mathbb E}$ stands for the expectation of the random variables $(g_x)_{z \in \Lambda_L}$.

Next, we consider 
replica symmetry breaking phenomena 
which apparently violate self-averaging 
of the overlap between two replicated quantities in a replica symmetric expectation.   
Let  $\phi^a (a=1, \cdots, n)$ be $n$ replicated copies of a spin configuration, and we consider the following Hamiltonian 
$$
H(\phi^1, \cdots, \phi^n, g):= \sum_{a=1}^n H(\phi^a, g),
$$
where replicated spin configurations share the same quenched disorder $g$.
This Hamiltonian is invariant under an arbitrary permutation $\sigma \in S_n$.
$$ H(\phi^1, \cdots, \phi^n, g)=H(\phi^{\sigma1}, \cdots, \phi^{\sigma n} , g)$$
 
This permutation symmetry is the replica symmetry.
The spin overlap $R_{a,b}$ between two replicated spin configurations is defined by   
$$
R_{a,b}:=\frac{1}{|\Lambda_L|}\sum_{x \in \Lambda_L} \phi_x^a \phi_x^b.
$$
The covariance of the Hamiltonian is written in terms of the overlap
$$
{\mathbb E} H(\phi^a,g) H(\phi^b,g) - {\mathbb E} H(\phi^a,g) {\mathbb E}H(\phi^b, g)=  |\Lambda_L| h^2R_{a,b}.
$$
When the replica symmetry breaking occurs, broadening of the overlap distribution with a finite variance is observed.
This phenomenon is  well-known in several disordered systems, such as the Sherrington-Kirkpatrick model \cite{Pr,T2,T}. 
In the present paper, we define replica symmetry breaking by the finite  variance  calculated in the replica symmetric expectation
in the infinite volume limit
$$
\lim_{L \rightarrow \infty} {\mathbb E}  \langle {\Delta R_{1,2} }^2 \rangle
 \neq 0,
$$
where
$\Delta R_{1,2} := R_{1,2} -{\mathbb E }  \langle R_{1,2} \rangle  $.  
Chatterjee has given this definition of the replica symmetry breaking and 
proved  
\begin{equation}
\lim_{L \rightarrow \infty} {\mathbb E}  \langle{ \Delta R_{1,2} }^2 \rangle
 =0,
 \label{0Var}
 \end{equation}
in the random field Ising model \cite{C2}. In the present paper, we extend his proof to 
the random field Ginzburg-Landau model.
We prove the following theorem.

{\theorem In the random field ferromagnetic Ginzburg-Landau model,  the following  variance vanishes \label{MT}
\begin{equation}
\lim_{L \rightarrow \infty} [
 {\mathbb E}  \langle {R_{1,2} }^2 \rangle - ({\mathbb E }  \langle R_{1,2} \rangle  )^2 ]=0,
\end{equation}
for almost all coupling constants 
$(\beta,u,h,r) \in [0,\infty)^3 \times {\mathbb R}^2$ in the infinite volume limit.
}
Here, we roughly sketch  Chatterjee's proof for the random field Ising model and  explain a key
to extend it to the random field Ginzburg-Landau model.

From the view point of detecting the spontaneous symmetry breaking, observe
another variance
$$
\lim_{L \rightarrow \infty}  \langle{ \delta R_{1,2}} ^2 \rangle,
$$
calculated in the replica symmetric Gibbs state,
where 
$\delta R_{1,2} :=R_{1,2}  -  \langle R_{1,2} \rangle.$ 
If this variance does not vanish in a sample, 
a strong fluctuation should yield an instability of the replica symmetric Gibbs state 
and one can expect spontaneous replica symmetry breaking. This phenomenon corresponds to a long range order in systems without disorder as discussed by Griffiths \cite{Gff}.  
First, Chatterjee prove 
 \begin{equation}
\lim_{L \rightarrow \infty} {\mathbb E}  \langle{ \delta R_{1,2} }^2 \rangle=0,
\label{0var}
\end{equation}
by the FKG inequality and another inequality obtained from the boundedness of $R_{1,2}$
in the random field Ising model.  In the random field Ginzburg-Landau model, however, there is no simple bound
for the truncated correlation function
$$
\langle \phi_x \phi_y \rangle -\langle \phi_x \rangle \langle \phi_y \rangle,
$$ 
unlike its sample expectation 
$$
{ \mathbb E}( \langle \phi_x \phi_y \rangle -\langle \phi_x \rangle \langle \phi_y \rangle)\leq C,
$$
for a positive number $C$. To show the limit (\ref{0var}) in the random field Ginzburg-Landau model, we prove a bound
$$
\sum_{x,y \in \Lambda_L}{ \mathbb E}( \langle \phi_x \phi_y \rangle -\langle \phi_x \rangle \langle \phi_y \rangle)^2 \leq K |\Lambda_L|^{-\frac{1}{4}},
$$ 
for a positive number $K$.  
Next, Chatterjee  points out a relation between two variances ${\mathbb E}\langle \Delta R_{1,2}^2 \rangle$ 
and ${\mathbb E} \langle \delta R_{1,2}^2 \rangle$.
For the disordered Ising systems, 
\begin{equation}
2\lim_{L\to\infty}{\mathbb E}  \langle {\Delta R_{1,2} }^2 \rangle
=3\lim_{L\to\infty}{\mathbb E}  \langle {\delta R_{1,2} }^2 \rangle
=6\lim_{L\to\infty}{\mathbb E}  \langle \Delta R_{1,2}  \rangle^2
\label{deltaDelta}
\end{equation}
are obtained from the Aizenman-Contucci  \cite{AC,CG} or the Ghirlanda-Guerra identities \cite{C2,C1,CG2,CG3,CMS,GG}.  Therefore, the identity (\ref{0var}) implies stronger identity (\ref{0Var}), then 
the proof has been completed for the random field Ising model.
The naive extension of this argument to the random field Ginzburg-Landau model
derives  a kind of the Ghirlanda-Guerra type identities  including the self-overlap
$R_{1,1}$ because of $\phi_x^2 \neq 1$, they are not useful to prove the identities (\ref{deltaDelta}).  
We use a new  method  
to obtain the variance bound on $R_{1,1}$, and then we  obtain
the Ghirlanda-Guerra identities as a useful form to prove the identities (\ref{deltaDelta}).
In the present paper, we prove that all variances of the spin overlap vanish in the replica symmetric Gibbs state
in another way for the random field ferromagnetic Ginzburg-Landau model.


\section{
Proof
}

\subsection{Properties of the free energy density in the infinite volume limit}

{\lemma \label{lem-bound-1point} The expectation of function of spin variable at a single site has an upper bound
\begin{equation}
{\mathbb E} \langle \phi_x^k \rangle  \leq C_k,
\label{bound-1point}
\end{equation}
where $k$ is an arbitrary even integer and $C_k$ is independent of the system size $L$. 
\\
Proof.} 
This is proved by an inductivity for even integer $k$.  First, we consider this bound in the case $k =2$.
Define the following interpolating function $\pi_L(s)$ of a parameter $s \in [0,1]$ for a positive number $b$ 
$$
\pi_L(s) :=\frac{1}{|\Lambda_L|} {\mathbb E} \log \int \prod_{x \in \Lambda_L} d\phi_x \exp [\sum_{x\in\Lambda_L} 
(-u \phi_x^4 + (b s +r) \phi_x^2 + \beta h g_x \phi_x)+\beta(1- s)
 \sum_{x,y \in \Lambda_L} J_{x,y} \phi_x \phi_y].
$$
Note that $\pi_L(0) = p_L(\beta, u, h,r)$ and $\pi_L(1)$ is defined in an independent spin model.
Since $\pi_L$ is convex in $s$, we have a bound 
\begin{eqnarray*}
 && \frac{1}{|\Lambda_L|}{\mathbb E}[ b \sum_{x \in \Lambda_L}  \langle \phi_x^2 \rangle 
- \beta  \sum_{x,y \in \Lambda_L} J_{x,y} \langle \phi_x \phi_y \rangle] 
= \pi_L' (0) \leq  \pi_L'(1)  \\  
&&=\frac{1}{|\Lambda_L|}{\mathbb E}[ b \sum_{x \in \Lambda_L}  \langle \phi_x^2 \rangle_0 
- \beta  \sum_{x,y \in \Lambda_L} J_{x,y} \langle \phi_x \phi_y \rangle_0].
  \label{bound-1point2,4}
\end{eqnarray*}
If we use  $2 \phi_x\phi_y \leq \phi_x^2+\phi_y^2$, the above inequality and the translational symmetry, we have
\begin{eqnarray*}
&&(b -\beta d) {\mathbb E} \langle \phi_x^2 \rangle\\
&&\leq\frac{1}{|\Lambda_L|}{\mathbb E}[b  \sum_{x \in \Lambda_L}  \langle \phi_x^2 \rangle
-  \frac{\beta}{2} \sum_{x,y \in \Lambda_L} J_{x,y}( \langle \phi_x^2 \rangle+ \langle \phi_y^2 \rangle)]\\
&&\leq \frac{1}{|\Lambda_L|}{\mathbb E}[b  \sum_{x \in \Lambda_L}  \langle \phi_x^2 \rangle_0 
+  \frac{\beta}{2} \sum_{x,y \in \Lambda_L} J_{x,y}( \langle \phi_x^2 \rangle_0+ \langle \phi_y^2 \rangle_0)]
 \\  &&\leq (b +\beta d) {\mathbb E} \langle \phi_x^2 \rangle_0, 
\end{eqnarray*}
Note that   
the right hand side in the above denotes the Gibbs expectation in the independent  spin model. Then the left hand side  ${\mathbb E} \langle
\phi_x^2 \rangle$
is  finite for a sufficiently large $b$, since the
one point function ${\mathbb E} \langle
\phi_x^2 \rangle_0$ in the independent model
has a simple upper bound.

Next, we consider the case for an even integer $k \geq 4$. 
An integration by parts with respect to a spin variable $\phi_x$ at a fixed site $x \in \Lambda_L$ gives
\begin{eqnarray}
&&(k-3) \langle  \phi_x^{k-4}\rangle =- \langle  \phi_x^{k-3}  \frac{\partial }{\partial \phi_x} 
 [ -u \phi_x^4+r\phi_x^2 +\sum_{y \in \Lambda_L}\beta J_{x,y} \phi_x \phi_y +\beta h g_x  \phi_x ] \rangle \nonumber  \\
&& = \langle 
 [ 4u \phi_x^{k}-2r\phi_x^{k-2} - \sum_{y \in \Lambda_L}\beta J_{x,y} \phi_x^{k-3} \phi_y  -\beta h g_x  \phi_x^{k-3})] \rangle\nonumber \\
 && = 
  4u \langle  \phi_x^{k} \rangle -2r \langle \phi_x^{k-2} \rangle - \sum_{y \in \Lambda_L}\beta J_{x,y} \langle  \phi_x^{k-3} \phi_y \rangle -\beta h g_x \langle  \phi_x^{k-3} \rangle.
  \label{intbypar1}
\end{eqnarray}
This equality (\ref{intbypar1}), H\"older's and the Jensen's  inequalities give the following recursive inequality for
 a bound on ${\mathbb E} \langle \phi_x^k \rangle$  for an even integer $ k>4$  
\begin{eqnarray*}
4u {\mathbb E}\langle  \phi_x^k \rangle
&=&(k-3){\mathbb E} \langle  \phi_x^{k-4} \rangle 
   +2r {\mathbb E}\langle \phi_x^{k-2} \rangle   +\beta h {\mathbb E} g_x \langle  \phi_x^{k-3} \rangle+ \sum_{y \in \Lambda_L}\beta J_{x,y} {\mathbb E}\langle  \phi_x^{k-3} \phi_y \rangle 
 \\
  &=& (k-3){\mathbb E} \langle  \phi_x^{k-4} \rangle 
   +2r {\mathbb E}\langle \phi_x^{k-2} \rangle   +\beta^2 h^2 {\mathbb E}( \langle  \phi_x^{k-2} \rangle
   -\langle  \phi_x^{k-3} \rangle \langle \phi_x \rangle)
 + \sum_{y \in \Lambda_L}\beta J_{x,y} {\mathbb E}\langle  \phi_x^{k-3} \phi_y \rangle 
 \\
&\leq&  (k-3){\mathbb E} \langle  \phi_x^{k-4} \rangle 
   +2r {\mathbb E}\langle \phi_x^{k-2} \rangle   +\beta^2 h^2 {\mathbb E} \langle  \phi_x^{k-2} \rangle 
 \\
&+& \beta^2 h^2 ({\mathbb E} |\langle  \phi_x^{k-3} \rangle|^\frac{k-2}{k-3} )^\frac{k-3}{k-2}
  ( {\mathbb E} \langle  \phi_x \rangle^{k-2})^\frac{1}{k-2}+ \sum_{y \in \Lambda_L}\beta J_{x,y} ({\mathbb E}\langle  \phi_x^{k-2} \rangle)^\frac{k-3}{k-2} ({\mathbb E} \langle \phi_y^{k-2} \rangle)^\frac{1}{k-2}\\
 &\leq&  (k-3){\mathbb E} \langle  \phi_x^{k-4} \rangle 
   +(2r 
 + 2\beta^2 h^2 + \sum_{y \in \Lambda_L}\beta J_{x,y}) {\mathbb E}\langle  \phi_x^{k-2} \rangle.
\end{eqnarray*}
Therefore, ${\mathbb E} \langle \phi_x^k \rangle$ for an even integer $k$ is bounded from the above by  ${\mathbb E} \langle \phi_x^{k-2} \rangle$. 
This and the bound  (\ref{bound-1point2,4}) for $k=2$ complete the proof.
$\Box$\\

{\lemma \label{existp}
 The following  infinite volume limit exists 
$$
p(\beta,h,u,r) = \lim_{L \rightarrow \infty} p_L(\beta,h,u,r), 
$$
for arbitrary coupling constants.
\\
Proof.} 
This is proved by a standard argument based on the decomposition of the lattice  into disjoint blocks \cite{CGP,CL,Gff}. 
 Let $L, M$ be positive integers and denote  $N=LM$, then we divide the lattice $\Lambda_N$ into $M^d$ disjoint translated blocks of 
 $\Lambda_L$.
 Define a new Hamiltonian $H$ on $\Lambda_N$ by deleting interaction bonds
 near the boundaries of blocks, 
 such that $M^d$ spin systems on the block $\Lambda_L$ and its $M^d-1$ translations have no  interaction with each other. 
 The original Hamiltonian $H_N$ has the following two terms
 $$
 H_N(\phi,g)=H(\phi,g)+H_{\rm del}(\phi,g),
 $$
 where $H_{\rm del}$ is deleted  interaction Hamiltonian, and $H$ denotes the  summation of Hamiltonians on the $M^d$ disjoint blocks. 
 Define the following function of $s \in [0,1]$ by
 $$
 \pi_N(s) :=\frac{1}{|\Lambda_N|} {\mathbb E}\log \int D\phi \exp[\beta (H+ s H_{\rm del})].
 $$
 Note that $\pi_N(1)=p_N(\beta,u,h,r)$ and $\pi_N(0)= p_L(\beta,u,h,r).$ 
 The derivative functions of $\pi_N$ are given by
 $$
 \pi_N'(s) =\frac{\beta}{|\Lambda_N|}  {\mathbb E} \langle H_{\rm del} \rangle_s, \ \ \ 
 \pi_N''(s) =\frac{\beta^2 }{|\Lambda_N|}  {\mathbb E} \langle (H_{\rm del}-  \langle H_{\rm del} \rangle_s )^2\rangle_s \geq 0,
  $$
 where 
 $\langle f(\phi)\rangle_s$ is the Gibbs and expectations  
  with the Hamiltonian $H+s H_{\rm del}$ for function  $f$ of a spin configuration.  
  Since the function $\pi_N(s)$ is convex, we have
 $$
 \pi_N' (0) \leq \pi_N(1) - \pi_N(0) \leq \pi'_N(1),
 $$
 then
 $$
  \frac{\beta}{N^d}  {\mathbb E} \langle H_{\rm del} \rangle_0 
\leq  p_N(\beta, u,h,r) - p_L(\beta, u,h,r) \leq \frac{\beta}{N^d}  {\mathbb E}  \langle H_{\rm del} \rangle_1
 $$
 Since the expectation 
 $$
 |{\mathbb E} \langle \phi_x \phi_y \rangle_s |  \leq \frac{1}{2}  ({\mathbb E} \langle \phi_x^2\rangle_s +{\mathbb E} \langle  \phi_y^2 \rangle_s),
 $$ is bounded as shown in Lemma \ref{lem-bound-1point},  $|{\mathbb E}  \langle H_{\rm del} \rangle_s|$  is bounded by the number of deleted bonds. Then, 
there exist  positive numbers $K$ independent of $L$ and $N$, such that the function $p_N$ and $p_L$ obey  
$$
|p_N-p_L| \leq\frac{K N^{d-1} (M-1)d}{N^d} \leq \frac{Kd}{L}.
$$
In the same argument for $M$ instead of $L$, we have
$$
|p_{N}-p_M|  \leq \frac{Kd}{M},
$$
and therefore 
$$
|p_L-p_{M}| \leq |p_N-p_L| +|p_{M}-p_N| \leq \frac{Kd}{L}+\frac{Kd}{M},
$$
The sequence $p_L$ is Cauchy.
$\Box$

{\note
The functions  $p$, $p_L$ $\psi_L$ are convex functions of each argument for arbitrarily  fixed others. }

{\note The function $p$ is differentiable almost everywhere in the coupling constant space $[0,\infty)^3
\times {\mathbb R}$ because of its convexity. }

Hereafter, we use a lighter notation $\psi_L(g)$ for $\psi_L(\beta,h,u,r,g)$.  
Define the function of $s \in [0,1]$
$$
G(s) :=  \sqrt{s} g_x +\sqrt{1-s} {g_x }' ,
$$
where  $g=(g_x)_{x \in \Lambda_L}$ and $g'=(g'_x)_{x \in \Lambda_L}$
consist of i.i.d. standard Gaussian variables. 
Define a generating  function $\gamma_L(s)$ of a parameter $s \in [0,1] $ by
\begin{equation}
\gamma_L(s) = {\mathbb E} [{\mathbb E}' \psi_L(G(s))]^2,
\end{equation}
where ${\mathbb E }$ and  ${\mathbb E }'$ denote expectation
over $g$ and $g'$, respectively. This generating function $\gamma_L$ is  introduced by Chatterjee \cite{C}.
We denote the Gibbs expectation of a function of spin configuration $f(\phi)$ with  Hamiltonian $H(\phi, G(s))$ 
\begin{equation}
\langle f(\phi) \rangle_{G(s)} := \frac{1}{Z_L(\beta,h,u,r, G(s))} \int D \phi f(\phi) \exp \Big[ -\beta H(\phi, G(s)) \Big].
\label{expectG}
\end{equation}
{\lemma \label{1}  
For any $(\beta,h,u,r) \in [0,\infty)^3 \times {\mathbb R}$, any positive integer
$L$, any positive integer $k$ and any $s_0 \in [0,1]$, 
an upper bound on the $k$-th order partial derivative of the function $\gamma_L$ is given by 
\begin {equation}
\frac{\partial^k \gamma_L}{{\partial s}^k} (s_0) \leq \frac{(k-1)! }{(1-s_0)^{k-1}}\frac{\beta^2 h^2 C_2 }{|\Lambda_L|}.
\label{kthi}
\end{equation}
The $k$-th order derivative of $\gamma_L$ is represented in the following
\begin{eqnarray}
\frac{\partial^k \gamma_L}{{\partial s}^k} (s) &=&
\sum_{x_1 \in \Lambda_L}  \cdots \sum_{x_k \in \Lambda_L}
{\mathbb E} \left({\mathbb E}' \frac{\partial^k \psi_L }{\partial G_{x_k} \cdots \partial G_{x_1} }(G(s)) \right)^2 .
\label{kth}
\end{eqnarray}
for an arbitrary $s \in [0,1].$ \\
\noindent Proof. 
} 
The first derivative  of $\gamma_L$  is calculated in integration by parts
\begin{eqnarray*}
&&\gamma_L' (s) = \frac{1}{|\Lambda_L|} {\mathbb E} \Big[{\mathbb E}'  \psi_L (G(s))
{\mathbb E}' \sum_{x \in \Lambda_L} \Big( \frac{g_x}{\sqrt{s}} -\frac{g'_x}{\sqrt{1-s}}  \Big) \frac{\partial \psi_L }{ \partial G_{x} }(G(s)) \Big]\\
&&= \frac{1}{|\Lambda_L|} {\mathbb E}\sum_{x \in \Lambda_L}\Big[  \frac{1}{\sqrt{s}}\frac{\partial}{\partial g_x}{\mathbb E}'   \psi_L (G(s))  {\mathbb E}' \frac{\partial \psi_L }{ \partial G_{x} }(G(s)) 
-{\mathbb E}'   \psi_L (G(s)) {\mathbb E}'   \frac{1}{\sqrt{1-s}}\frac{\partial}{\partial g'_x}  \frac{\partial \psi_L }{ \partial G_{x} }(G(s)) \Big]\\
&& = \sum_{x \in \Lambda_L} 
{\mathbb E} \left({\mathbb E}' \frac{\partial \psi_L }{ \partial G_{x} }(G(s)) \right)^2=\frac{\beta^2 h^2}{|\Lambda_L|^2}  \sum_{x \in \Lambda_L} 
 {\mathbb E} \Big({\mathbb E}'   \langle \phi_x  \rangle_{G(s)}  \Big)^2 \leq\frac{\beta^2 h^2}{|\Lambda_L|^2}  \sum_{x \in \Lambda_L} 
 {\mathbb E} \langle \phi_x  \rangle ^2 \leq \frac{\beta^2 h^2 C_2}{|\Lambda_L|}.
 \end{eqnarray*}
 The bound ${\mathbb E}\langle \phi_x^2 \rangle \leq C_2$ given by  Lemma \ref{lem-bound-1point}  has been used.
The formula (\ref{kth}) for the $k$-th derivative is proved by inductivity.
The positive semi-definiteness of arbitrary order derivative $\gamma_L^{(k)} (s)$ and
Taylor's theorem 
$$
\gamma_L'(1) \geq  \sum_{j=0} ^{k-2} \frac{(1-s_0)^j}{j!} \gamma_L^{(j+1)} (s_0)  + \frac{(1-s_0)^{k-1}}{(k-1)!} \gamma_L^{(k)} (s_1) 
$$
for $s_1 \in (s_0,1)$ give the first inequality (\ref{kthi}). $\Box$\\

Lemma \ref{1} gives the following lemma

{\lemma \label{free}  The variance of 
$ \psi_L$ is bounded from the above as follows
$$
  Var (\psi_L) \leq   \frac{\beta^2 h^2 C_2 }{|\Lambda_L|}.
$$
\noindent
Proof.
} The left hand side is given by
$$
{\mathbb  E} (\psi_L-p_L )^2 =\gamma_L(1)-\gamma_L(0)  =\int_0 ^1 ds \gamma_L'(s) \leq \gamma_L'(1) \leq \frac{\beta^2 h^2C_2}{|\Lambda_L|}.
$$
This completes the proof. $\Box$

\subsection{Variance inequalities for the Hamiltonian density}

Next we evaluate the following variance  of the random Hamiltonian density defined by
$$\xi_L :=\frac{1}{|\Lambda_L|} \sum_{x\in \Lambda_L} g_x \phi_x
$$
{\lemma  \label{deltah} 
For any coupling constants,  we have 
\begin{equation}
 {\mathbb E}\langle \delta { \xi_L}^2\rangle \leq  \frac{1}{\beta h} \sqrt{\frac{6C_2}{|\Lambda_L|}} + \frac{C_2}{|\Lambda_L|} , \label{a1}
\end{equation}
where $\delta \xi_L:= \xi_L- \langle \xi_L \rangle$.\\
Proof. 
} 
This bound on the variance  ${\mathbb E} \langle \delta \xi_L^2 \rangle $ is obtained as follows: 
\begin{eqnarray*}
&&{\mathbb E} \langle \delta \xi_L^2\rangle =\
\frac{1}{|\Lambda_L|^2} \sum_{x,y \in \Lambda}{ \mathbb E} g_x g_y (\langle \phi_x \phi_y \rangle-\langle \phi_x \rangle \langle\phi_y \rangle)\\
&&= \frac{1}{|\Lambda_L| ^2} \sum_{x,y \in \Lambda_L}{\mathbb E} \Big( \frac{\partial^2}{\partial g_x \partial g_y} +\delta_{x,y}\Big) 
(\langle \phi_x \phi_y \rangle-\langle \phi_x \rangle \langle\phi_y \rangle) \\
&&\leq\frac{1}{\beta^2 h^2 |\Lambda_L| } \sum_{x,y \in \Lambda_L}{\mathbb E} \frac{\partial^4 \psi_L}{\partial g_x ^2\partial g_y^2} + \frac{1}{|\Lambda_L|^2}
\sum_{x\in\Lambda_L}   (\langle \phi_x^2\rangle-\langle \phi_x \rangle^2) \\
&&\leq \frac{1}{\beta^2 h^2 |\Lambda_L| }\sqrt{\sum_{x,y \in \Lambda_L} \Big({ \mathbb E} \frac{\partial^4 \psi_L}{\partial g_x ^2\partial g_y^2} \Big)^2 \sum_{x,y \in \Lambda_L} 1^2 }
+\frac{C_2}{|\Lambda_L|}\\
&& \leq \frac{\sqrt{\gamma_L''''(0)}}{\beta^2h^2}  +\frac{C_2}{|\Lambda_L|} \leq \frac{1}{\beta h} \sqrt{\frac{6C_2}{|\Lambda_L|}} + \frac{C_2}{|\Lambda_L|}.
\end{eqnarray*} 
The boundedness of  the spins and  Lemma \ref{1}  have been used. $\Box$

Next lemma can be proved by a standard argument of  the continuous differentiability of the function  $p(\beta, h )$ for almost all  $h$ because of the convexity \cite{C2,T}.

{\lemma  \label{Delta}  For almost all coupling constants,  
\begin{equation}
\frac{\partial p}{\partial h}=\lim_{L \rightarrow \infty}{\mathbb E} \frac{\partial \psi_L}{\partial h}, \label{lim1}
\end{equation}
and
\begin{equation}
\lim_{L\to\infty} {\mathbb E} (\langle \xi_L \rangle -{\mathbb E} \langle \xi_L \rangle )^2=0.
\end{equation}
Proof.
}
Regard $p_L(h)$ $p(h)$ and $\psi_L(h)$ as functions of $h$ for lighter notation. 
Define the following functions 
\begin{eqnarray}
w_L(\epsilon) &:=& \frac{1}{\epsilon}[|\psi_L(h+\epsilon )-p_L(h+\epsilon)|+|\psi_L(h- \epsilon)-p_L(h-\epsilon)|
+|\psi_L(h )-p_L(h)| ]\nonumber \\
e_L(\epsilon )&:=&\frac{1}{\epsilon}[|p_L(h+\epsilon )-p(h+\epsilon)|+|p_L(h- \epsilon)-p(h-\epsilon)|
+|p_L(h )-p(h)|],\nonumber
\end{eqnarray}
for $\epsilon > 0$.
Lemma \ref{existp}   gives
\begin{equation}
  \lim_{L \rightarrow\infty} e_L(\epsilon)=0,
\end{equation}
and  Lemma \ref{free} and the Schwarz inequality give  also
 \begin{equation}
\lim_{L \rightarrow\infty} {\mathbb E}w_L(\epsilon)=0, \ \ \ \lim_{L \rightarrow\infty} {\mathbb E}w_L(\epsilon)^2=0, 
\end{equation}
for any $\epsilon > 0$. 
Since $\psi_L$, $p_L$ and $p$ are convex functions of $h$, we have
\begin{eqnarray} 
&&\frac{\partial \psi_L}{\partial h}(h) - \frac{\partial  p}{\partial h}(h)
\leq \frac{1}{\epsilon} [\psi_L(h+\epsilon)-\psi_L(h)]- \frac{\partial  p}{\partial h}\nonumber \\
&&\leq \frac{1}{\epsilon} [\psi_L(h+\epsilon)-p_L(h+\epsilon)+p_L(h+\epsilon)-p_L(h)
+p_L(h)-\psi_L(h) \nonumber \\
&& -p(h+\epsilon) +p(h+\epsilon)+p(h)-p(h) ]- \frac{\partial  p}{\partial h}(h) \nonumber \\
&&\leq \frac{1}{\epsilon} [ |\psi_L(h+\epsilon)-p_L(h+\epsilon)|
+|p_L(h)-\psi_L(h)| +|p_L(h+\epsilon)-p(h+\epsilon)|\nonumber \\ 
&&+|p_L(h)-p(h)| ]+\frac{1}{\epsilon}[ p(h+\epsilon)-p(h)] - \frac{\partial  p}{\partial h}(h) \nonumber \\
&&\leq w_L(\epsilon) +e_L(\epsilon)
+  \frac{\partial p}{\partial h}(h+\epsilon) - \frac{\partial  p}{\partial h}(h). \nonumber
\end{eqnarray}  
As in the same calculation, we have
\begin{eqnarray} 
&&\frac{\partial \psi_L}{\partial bh}(h) - \frac{\partial  p}{\partial h}(h) 
\geq \frac{1}{\epsilon}[\psi_L(h)-\psi_L(h-\epsilon)] - \frac{\partial  p}{\partial h}(h) \nonumber \\&&
\geq -w_L(\epsilon) -e_L(\epsilon)+ \frac{\partial p}{\partial h}(h-\epsilon)- \frac{\partial  p}{\partial h}(h) . \nonumber 
\end{eqnarray}  
Then, 
\begin{eqnarray} 
\Big|\frac{\partial \psi_L}{\partial h}(h) - \frac{\partial p}{\partial h}(h)\Big| \leq w_L(\epsilon)
+e_L(\epsilon)+  \frac{\partial p}{\partial h}(h+\epsilon) -  \frac{\partial p}{\partial h}(h-\epsilon).\nonumber
\end{eqnarray}  
Convergence of ${\mathbb E} w_L,$ ${\mathbb E} w_L^2$ and $e_L$ in  the infinite volume limit implies 
\begin{eqnarray}
&&\lim_{L \rightarrow\infty } {\mathbb E}\Big| \beta \langle \xi_L  \rangle - \frac{\partial p}{\partial h}(h)\Big|^2
\leq  \Big[\frac{\partial p}{\partial h}(h+\epsilon) -  \frac{\partial p}{\partial h}(h-\epsilon) \Big]^2, \nonumber
\end{eqnarray}
The right hand side vanishes, since the convex function $p(h)$ is continuously 
differentiable almost everywhere and $\epsilon >0$ is arbitrary. Therefore
\begin{equation}
\lim_{L \rightarrow\infty}{\mathbb E} \Big| \beta  \langle \xi_L  \rangle - \frac{\partial p}{\partial h}(h)\Big|^2=0.
\label{limit3}
\end{equation}
for almost all $h$.   Jensen's inequality gives 
\begin{equation}
\lim_{L \rightarrow\infty}\Big| {\mathbb E}\beta  \langle \xi_L  \rangle - \frac{\partial p}{\partial h}(h)\Big|^2=0.
\end{equation}
 This implies the first equality (\ref{lim1}).  
These equalities imply also
$$
\lim_{L \rightarrow\infty }  {\mathbb E} (\langle \xi_L  \rangle - 
 {\mathbb E} \langle \xi_L  \rangle)^2=0.
$$
This  completes the proof.
$\Box$\\

{\lemma The following  limit vanishes
$$
\lim_{L \rightarrow \infty} {\mathbb E} \langle  {\Delta \xi_L} ^2\rangle=0,
$$
almost everywhere in coupling constant space, where $\Delta \xi_L := \xi_L - {\mathbb E} \langle \xi_L  \rangle$ \\
Proof.}
\begin{eqnarray*}
&&{\mathbb E} \langle  {\Delta \xi_L} ^2\rangle = {\mathbb E} \langle  \xi_L^2\rangle - {\mathbb E} \langle \xi_L  \rangle^2
+ {\mathbb E}\langle \xi_L  \rangle^2- ({\mathbb E} \langle \xi_L\rangle  )^2
\end{eqnarray*}
Lemma \ref{deltah} and  Lemma \ref{Delta} complete the proof. $\Box$

\subsection{Variance inequalities for spin overlap functions}


\noindent
{\bf Notation} {\it  An expression  $\langle A \ ;B \rangle$ denotes a two point truncated correlation function
$$\langle A \ ;B \rangle := \langle AB \rangle -\langle A\rangle \langle B \rangle.$$
for functions $A, B$ of spin configurations $(\phi_x)_{x\in \Lambda_L}$,
}

{\lemma  \label{FKG} 
Let $f$ and $g$ be monotonically increasing functions of spin configuration, then
$
\langle f ; g \rangle \geq 0.
$
 Therefore,   
 $
\langle \phi_x ; \phi_y \rangle \geq 0,
$
 for any $x,y \in \Lambda_L.$
}

This inequality   proved by Fortuin, Kasteleyn and  Ginibre  is called the FKG inequality \cite{FKG}.

{\lemma \label{bound-sq-2point} There exists a positive number $K$ independent of the system size $L$, such that 
the expectation of square of truncated two point function is bounded from the above 
$$
\frac{1}{|\Lambda_L|^2}\sum_{x,y\in \Lambda_L} {\mathbb E} \langle \phi_x; \phi_y\rangle^2 \leq K|\Lambda_L|^{-\frac{1}{4}}.
$$
Proof.}  
Define an indicator defined $I$ by $I[true]=1,$ and $I[false]=0$.
 Let $C$ be a positive number and evaluate
\begin{eqnarray*}
&&\frac{1}{|\Lambda_L|^2}\sum_{x,y\in \Lambda_L} {\mathbb E} I [\langle \phi_x; \phi_y\rangle \geq  C]
\leq\frac{1}{C|\Lambda_L|^2}\sum_{x,y\in \Lambda_L} {\mathbb E} \langle \phi_x; \phi_y\rangle  I [\langle \phi_x; \phi_y\rangle \geq  C]\\
&& \leq \frac{1}{C|\Lambda_L|^2}\sum_{x,y\in \Lambda_L} {\mathbb E} \langle \phi_x; \phi_y\rangle \\
&& \leq \frac{1}{C|\Lambda_L|^2}\sqrt{\sum_{x,y\in \Lambda_L} \Big( {\mathbb E} \langle \phi_x; \phi_y\rangle \Big)^2  \sum_{x,y\in \Lambda_L} 1^2}
\leq \frac{1}{C\beta h} \sqrt{\frac{C_2}{|\Lambda_L|}}.
\end{eqnarray*}
The FKG inequality  and Lemma \ref{1} have been used. Next we evaluate expectation of summation of squared two point correlation functions over
the lattice 
\begin{eqnarray*}
&&\frac{1}{|\Lambda_L|^2}\sum_{x,y\in \Lambda_L} {\mathbb E} \langle \phi_x; \phi_y\rangle^2 \\
&&\leq\frac{1}{|\Lambda_L|^2}\sum_{x,y\in \Lambda_L} {\mathbb E} \langle \phi_x; \phi_y\rangle^2 ( I [\langle \phi_x; \phi_y\rangle <  C]+ I [\langle \phi_x; \phi_y\rangle \geq  C]) \\
&&\leq\frac{1}{|\Lambda_L|^2}\sum_{x,y\in \Lambda_L} {\mathbb E}[ \langle \phi_x; \phi_y\rangle  C I [\langle \phi_x; \phi_y\rangle <  C]+
\langle \phi_x; \phi_y\rangle ^2 I [\langle \phi_x; \phi_y\rangle \geq  C]) \\
&&\leq\frac{1}{|\Lambda_L|^2}\sum_{x,y\in \Lambda_L} {\mathbb E}(\langle \phi_x; \phi_y\rangle  C+
\langle \phi_x; \phi_y\rangle ^2 I [\langle \phi_x; \phi_y\rangle \geq  C]) \\
&& \leq \frac{C}{|\Lambda_L|^2}\sqrt{\sum_{x,y\in \Lambda_L} \Big( {\mathbb E} \langle \phi_x; \phi_y\rangle \Big)^2  \sum_{x,y\in \Lambda_L} 1^2}
+\frac{1}{|\Lambda_L|^2}\sum_{x,y\in \Lambda_L} \sqrt{{\mathbb E}
\langle \phi_x; \phi_y\rangle ^4   {\mathbb E} I [\langle \phi_x; \phi_y\rangle \geq  C] }\\
&&\leq \frac{C}{\beta h}\sqrt{\frac{C_2}{|\Lambda_L|}} +4 \sqrt{ \frac{C_8}{|\Lambda_L|^2}\sum_{x,y\in \Lambda_L}  {\mathbb E} I [\langle \phi_x; \phi_y\rangle \geq  C] }
\leq \frac{C}{\beta h}\sqrt{\frac{C_2}{|\Lambda_L|}} +4 \sqrt{\frac{ C_8}{C\beta h}} \Big(\frac{C_2}{|\Lambda_L|}\Big)^\frac{1}{4},
\end{eqnarray*}
where an upper bound ${\mathbb E}
\langle \phi_x; \phi_y\rangle ^4  \leq 16 C_8$ is guaranteed by the bound on one point functions (\ref{bound-1point}) and the Schwarz inequality.
This implies the upper bound. $\Box$

{\lemma \label{delta} For any coupling constants 
\begin{equation}
{\mathbb E}[\langle {R_{1,2} }^2 \rangle   -\langle R_{1,2} \rangle ^2] \leq 2 \sqrt{C_4K} |\Lambda_L|^{-\frac{1}{8}}.
\end{equation}
\noindent Proof.
} 
This can be proved using the FKG inequality $\langle \phi_x ; \phi_y \rangle \geq 0$ as proved for the random field Ising model \cite{C2}. 
\begin{eqnarray*}
&&{\mathbb E}[\langle {R_{1,2} }^2 \rangle   -\langle R_{1,2} \rangle ^2] = \frac{1}{|\Lambda_L|^2} \sum_{x,y\in \Lambda_L}{\mathbb E}
 (\langle  \phi_x \phi_y\rangle ^2-\langle  \phi_x  \rangle^2 \langle \phi_y\rangle ^2)\\
&& \leq \frac{1}{|\Lambda_L|^2} \sum_{x,y\in \Lambda_L}{\mathbb E} 
 |\langle  \phi_x \phi_y\rangle -\langle  \phi_x  \rangle \langle \phi_y\rangle ||\langle  \phi_x \phi_y\rangle+
 \langle  \phi_x  \rangle \langle \phi_y\rangle |\\
&& \leq \frac{1}{|\Lambda_L|^2} \sum_{x,y\in \Lambda_L}{\mathbb E} 
 \langle  \phi_x ; \phi_y\rangle ( \sqrt{\langle  \phi_x^2  \rangle \langle \phi_y^2\rangle} + | \langle  \phi_x  \rangle|| \langle \phi_y\rangle| )
\\
&& \leq 
  \frac{1}{|\Lambda_L|^2} \sum_{x,y\in \Lambda_L} \Big( \sqrt{{\mathbb E} 
 \langle  \phi_x ; \phi_y\rangle^2  } 
\sqrt{ {\mathbb E} \langle  \phi_x^2  \rangle \langle \phi_y^2\rangle}+
\sqrt{{\mathbb E} 
 \langle  \phi_x ; \phi_y\rangle^2  } \sqrt{ {\mathbb  E}  \langle  \phi_x  \rangle^2 \langle \phi_y\rangle^2 } \Big)
 \\
 && \leq 
  \frac{2}{|\Lambda_L|^2} \sum_{x,y\in \Lambda_L}\sqrt{{\mathbb E} 
 \langle  \phi_x ; \phi_y\rangle^2  } 
\sqrt{ {\mathbb E} \langle  \phi_x^2  \rangle \langle \phi_y^2\rangle}\leq 
  \frac{2}{|\Lambda_L|^2} \sum_{x,y\in \Lambda_L}\sqrt{{\mathbb E} 
 \langle  \phi_x ; \phi_y\rangle^2  } 
\Big( {\mathbb E} \langle  \phi_x^2  \rangle^2 {\mathbb E} \langle \phi_y^2\rangle^2 \Big)^\frac{1}{4}
 \\
&&\leq \frac{2}{|\Lambda_L|^2} 
\sqrt{\sum_{x,y\in \Lambda_L} {\mathbb E} 
 \langle  \phi_x ; \phi_y\rangle ^2
  \sum_{x,y\in \Lambda_L}  C_4} \\
 &&\leq 2 \sqrt{C_4K} |\Lambda_L|^{-\frac{1}{8}}.
\end{eqnarray*}
We have used Lemma  \ref{bound-sq-2point}.  $\Box$

The following two lemmas show that two kinds of variance of self-overlap $R_{1,1}$ vanish. 
Proving these two lemmas  is necessary to obtain the Ghirlanda-Guerra identities for the random field Ginzburg-Landau model, 
although they are automatically valid in the random field 
Ising model because of $R_{1,1} =1$.

{\lemma \label{delta11}
For all coupling constants,  
\begin{equation}
\lim_{L\to\infty}{\mathbb E}[\langle R_{1,1}^2 \rangle -{\mathbb E} \langle R_{1,1}\rangle^2]=0.
\end{equation}
Prrof.}
To prove this lemma we have to evaluate the upper bound of the following Gibbs expectation
$$
\frac{1}{|\Lambda_L|^2}\sum_{x,y \in \Lambda_L} {\mathbb E} \langle \phi_x ^2 \ ; \phi_y^2 \rangle.
$$
To obtain the bound, represent the following derivative function in terms of correlation functions

 $$
\frac{|\Lambda_L|}{(\beta h)^4}  \sum_{x,y\in \Lambda_L}\frac{\partial^4 \psi_L}{\partial g_x^2 \partial g_y^2}= \sum_{x,y\in \Lambda_L} [\langle \phi_x^2 ; \phi_y^2 \rangle
 -2\langle \phi_x ; \phi_y \rangle(\langle \phi_x  \phi_y \rangle-3\langle \phi_x\rangle \langle  \phi_y \rangle )
 -4 \langle \phi_x^2 ; \phi_y \rangle\langle \phi_y \rangle ].
$$
Consider the following function to estimate the first and the last terms
$$
\chi_L(t) := \frac{1}{|\Lambda_L|^2}\sum_{x \in \Lambda_L} {\mathbb E} ({\mathbb E}' \langle \phi_x^2 \rangle_{G(t)} )^2, 
$$
where $(G_x(t))_{x\in\Lambda_L}:= (\sqrt{t} g_x + \sqrt{1-t} g_x')_{x\in\Lambda_L}$ for $t \in[0,1]$. The $k$-th derivative of $\chi_L$ is
given by 
\begin{eqnarray*}
\chi_L^{(k)}(t) 
&=& \frac{1}{|\Lambda_L|^2}\sum_{x,y_1,\cdots , y_k \in\Lambda_L}  {\mathbb E}
\Big [{\mathbb E}' \frac{\partial ^k}{\partial G_{y_k} \cdots \partial G_{y_1}}\langle \phi_x^2 \rangle_{G(t)} 
 \Big]^2.
\end{eqnarray*}
Note that any order derivative is monotonically increasing in $t$. The second derivative function of $\chi_L$ is represented in the following summation of  three point correlation functions
which gives us the bound on the summation of  two point functions.
\begin{eqnarray*}
\chi_L''(t) 
&=& \frac{(\beta h)^4}{|\Lambda_L|^2}\sum_{x,y,z \in\Lambda_L}  {\mathbb E}
 [{\mathbb E}' (\langle \phi_x^2  \phi_y ;\phi_z\rangle_{G(t)}  - 
  \langle\phi_x^2  \ ; \phi_z  \rangle_{G(t)} \langle \phi_y  \rangle_{G(t)} - \langle\phi_y  \ ; \phi_z  \rangle_{G(t)} \langle \phi_x^2  \rangle_{G(t)} )]^2 \\
  &\geq& \frac{(\beta h)^4}{|\Lambda_L|^2}\sum_{x,y\in\Lambda_L}  {\mathbb E}[ {\mathbb E}'(\langle \phi_x^2 \ ; \phi_y^2 \rangle_{G(t)}
-2\langle\phi_x^2 \ ; \phi_y  \rangle_{G(t)} \langle \phi_y  \rangle_{G(t)} )]^2.
\end{eqnarray*}
A bound for $\chi_L''(0)$ is given in terms of  $\chi_L(1)$ using Taylor's theorem as for $\gamma_L^{(k)}(0)$
\begin{eqnarray*}
&&\frac{C_2^2}{|\Lambda_L|}\geq \chi_L(1) \geq \frac{1}{2} \chi_L''(0) 
= \frac{(\beta h)^4}{2|\Lambda_L|^2}\sum_{x,y,z \in\Lambda_L}  [{\mathbb E} (\langle \phi_x^2 \phi_y ; \phi_z \rangle - 
 \langle\phi_x^2  \ ; \phi_z  \rangle\langle \phi_y  \rangle - \langle\phi_y  \ ; \phi_z  \rangle\langle \phi_x^2  \rangle )]^2\\
&&\geq \frac{(\beta h)^4}{2|\Lambda_L|^2}\sum_{x,y\in\Lambda_L}  ({\mathbb E}\langle \phi_x^2 \ ; \phi_y^2 \rangle
-2 {\mathbb E}\langle\phi_x^2 \ ; \phi_y  \rangle \langle \phi_y  \rangle )^2.
\end{eqnarray*}
The following correlation is estimated using the above bound
\begin{eqnarray}
\sum_{x,y\in \Lambda_L} {\mathbb E} \langle \phi_x^2 ; \phi_y^2 \rangle
&=& \sum_{x,y\in \Lambda_L}{\mathbb E} 
[2 \langle \phi_x^2 ; \phi_y^2 \rangle-4 \langle \phi_x^2 ; \phi_y \rangle\langle \phi_y \rangle-2\langle \phi_x ; \phi_y \rangle(\langle \phi_x  \phi_y \rangle-3\langle \phi_x\rangle \langle  \phi_y \rangle )
  ]\nonumber \\
&-&\frac{|\Lambda_L|}{(\beta h)^4}  \sum_{x,y\in \Lambda_L}{\mathbb E} \frac{\partial^4 \psi_L}{\partial g_x^2 \partial g_y^2}.
\end{eqnarray}
 We obtain the following bound 
\begin{eqnarray*}
&& \Big| \sum_{x,y\in \Lambda_L}{\mathbb E} \langle \phi_x^2 ; \phi_y ^2\rangle  \Big|\leq 
  2   \sum_{x,y\in \Lambda_L}\Big|{\mathbb E}(\langle \phi_x^2 ; \phi_y^2 \rangle-2  \langle \phi_x^2 ; \phi_y \rangle\langle \phi_y \rangle) \Big| \\
&& +2 \sum_{x,y\in \Lambda_L}\Big|{\mathbb E} \langle \phi_x ; \phi_y \rangle(\langle \phi_x  \phi_y \rangle-3\langle \phi_x\rangle \langle  \phi_y \rangle )\Big|
+ \frac{|\Lambda_L|}{(\beta h)^4}  \sum_{x,y\in \Lambda_L}\Big| {\mathbb E} \frac{\partial^4 \psi_L}{\partial g_x^2 \partial g_y^2} \Big|  \\
 && \leq  2\sqrt{ \sum_{x,y\in \Lambda_L} \Big(  {\mathbb E}(\langle \phi_x^2 ; \phi_y ^2\rangle -
2  \langle \phi_x^2 ; \phi_y \rangle\langle \phi_y \rangle)\Big) ^2
 \sum_{x,y \in \Lambda_L} 1^2}  \\
&&+2\sum_{x,y\in \Lambda_L} \sqrt{   {\mathbb E}\langle \phi_x; \phi_y\rangle ^2}
 ( \sqrt{ {\mathbb E} \langle \phi_x ^2\rangle \langle \phi_y^2\rangle } + 3  \sqrt{ {\mathbb E} \langle \phi_x \rangle^2 \langle \phi_y\rangle^2 } )
 +\frac{|\Lambda_L|}{(\beta h)^4}\sqrt{ \sum_{x,y\in \Lambda_L} \Big(  {\mathbb E} \frac{\partial^4 \psi_L}{\partial g_x^2 \partial g_y^2}\Big) ^2
 \sum_{x,y \in \Lambda_L} 1^2} \\
 &&\leq \frac{2}{(\beta h)^2}\sqrt{ \chi_L''(0) |\Lambda_L|^4 }+
8  \sqrt{C_4K}
 | \Lambda_L|^\frac{15}{8}  +\frac{|\Lambda_L|}{(\beta h)^4}\sqrt{ \gamma_L''''(0) | \Lambda_L|^2} \\
 &&\leq 8 \sqrt{C_4 K}
 | \Lambda_L|^\frac{15}{8}  + \left(\frac{2\sqrt{2}C_2}{\beta^2 h^2}+ \frac{\sqrt{6C_2}}{\beta^3 h^3} \right)|\Lambda_L|^\frac{3}{2}.
 \end{eqnarray*}
 The FKG inequality for $\langle \phi_x ; \phi_y \rangle$, bounds on $\chi_L''(0)$, $\gamma_L''''(0)$ 
 and Lemma \ref{bound-sq-2point} have been used.  These estimates conclude that  
$$\lim_{L \to \infty} 
\frac{1}{|\Lambda_L|^2}\sum_{x,y \in \Lambda_L} {\mathbb E} \langle \phi_x ^2 \ ; \phi_y^2 \rangle =0.
$$
Then the variance  of $R_{1,1} $ vanishes in the infinite volume limit. $\Box$\\

Next lemma can be proved by the continuous differentiability of the function  $p(\beta, r )$  in $r$ for almost all  $r$ because of its convexity 
as in the proof of Lemma \ref{Delta}.

{\lemma \label{Delta11}
 For almost all coupling constants,  
\begin{equation}
\frac{\partial p}{\partial r}=\lim_{L \rightarrow \infty}{\mathbb E} \frac{\partial \psi_L}{\partial r},\label{lim11}
\end{equation}
and
\begin{equation}
\lim_{L\to\infty} {\mathbb E} (\langle R_{1,1} \rangle -{\mathbb E} \langle R_{1,1} \rangle )^2=0.
\end{equation}
}

\subsection{The Ghirlanda-Guerra identities}

 Lemma \ref{Delta}, \ref{delta} and \ref{Delta11}   enable us to derive the  Ghirlanda-Guerra identities for the Ginzburg-Landau model
in the useful form as well as  those for  Ising systems  \cite{AC,GG} .

{\lemma  \label{MT2}
Let  $f(\phi^1, \cdots, \phi^n)$ be an arbitrary function of $n$ replicated spins, satisfying  a bound ${\mathbb E }\langle f^2 \rangle \leq C_f^2$.
For $\beta h \neq 0$,  almost everywhere in the coupling constant space, 
the following identity is valid
\begin{eqnarray}
\lim_{L \rightarrow \infty} \Big[\sum_{a=2} ^n {\mathbb E}\langle R_{1,a} f\rangle -n {\mathbb E} \langle R_{1,n+1} f \rangle
 \ + \ {\mathbb E}\langle R_{1,2} \rangle {\mathbb E}\langle f \rangle
\Big]=0.  
\label{GG}
\end{eqnarray}
Proof. } 
 Lemma \ref{Delta},  the boundedness of $f$ and the Schwarz inequality  imply 
$$
|{\mathbb E}\langle  \Delta \xi_L f\rangle|  \leq \sqrt{{\mathbb E}\langle  {\Delta \xi_L}^2  \rangle  
{\mathbb E}\langle f^2 \rangle} \leq \sqrt{{\mathbb E}\langle  {\Delta \xi_L}^2  \rangle  C_f^2 } \rightarrow 0,
$$ 
in the infinite volume limit.
The left hand side can be calculated using integration by parts.
\begin{eqnarray}
&&\frac{1}{|\Lambda_L|}\sum_{x \in \Lambda_L}{\mathbb E}  g_x \langle   \phi_x^1 f\rangle 
=\frac{1}{|\Lambda_L|} \sum_{x \in \Lambda_L} {\mathbb E} \frac{\partial }{\partial g_x} \langle \phi_x^1 f\rangle \nonumber \\
&&= \frac{\beta h}{|\Lambda_L|} \sum_{x \in \Lambda_L} [\sum_{a=1}^n {\mathbb E}\langle \phi_x^1, \phi_x^{a} f \rangle -n  {\mathbb E}\langle  \phi_x \rangle  \langle   \phi^1_x f \rangle ]  \nonumber \\
&&=\beta h \Big[\frac{1}{|\Lambda_L|} \sum_{x \in \Lambda_L} \sum_{a=1}^n {\mathbb E}\langle \phi_x^a\phi_x^1 f\rangle-n {\mathbb E}\langle R_{1,n+1} f \rangle \Big] 
\end{eqnarray}
Substituting  $f=1$ to the above,  we have
\begin{eqnarray}
&&\frac{1}{|\Lambda_L|}\sum_{X \in \Lambda_L}{\mathbb E}  g_x \langle \phi_x \rangle 
\nonumber \\
&&=\beta h \Big[\frac{1}{|\Lambda_L|} \sum_{x \in \Lambda_L}{ \mathbb E}\langle \phi_x^2\rangle+
 \sum_{\alpha=2}^n { \mathbb E}\langle R_{1,a }\rangle-n { \mathbb E}\langle R_{1,n+1} \rangle \Big] \nonumber \\
&&=\beta h[{ \mathbb E}\langle R_{1,1} \rangle - { \mathbb E}\langle R_{1,2} \rangle ]
\end{eqnarray}
From the above two identities, we have 
\begin{eqnarray}
{\mathbb E} \langle \Delta \xi_L f \rangle
&= &\beta h \Big[  \sum_{a=1}^n {\mathbb E}\langle R_{1,a} f \rangle
-n {\mathbb E}\langle R_{1,n+1} f  \rangle
-({ \mathbb E}\langle R_{1,1} \rangle - { \mathbb E}\langle R_{1,2} \rangle) {\mathbb E } \langle f \rangle \Big]
\nonumber \\
&=& \beta h \Big[  \sum_{a=2}^n {\mathbb E}\langle R_{1,a} f \rangle
-n {\mathbb E}\langle R_{1,n+1} f  \rangle
+ { \mathbb E}\langle R_{1,2} \rangle{\mathbb E } \langle f \rangle
+{\mathbb E} \langle  \Delta R_{1,1} f \rangle \Big],
\end{eqnarray}
Therefore,  Lemma \ref{Delta11} and Lemma \ref{delta11} enable us to obtain the identity (\ref{GG}) $\Box$ \\

\subsection{Concluding  the proof of Theorem \ref{MT}}

As proved by Chatterjee for the random field Ising model  \cite{C2}, we use the Ghirlanda-Guerra identities.
Lemma \ref{MT2} for $n=2$ and  $f=R_{1,2} $
implies 
$$\lim_{L\to\infty}[2{\mathbb E}\langle R_{1,2} R_{1,3}\rangle-{\mathbb E}\langle R_{1,2}^2\rangle -({\mathbb E}\langle R_{1,2}\rangle)^2]=0,
$$
and it for  $n=3$ and $  f=R_{2,3}$ implies
$$
\lim_{L\to\infty}[3{\mathbb E}\langle R_{2,3} R_{1,4}\rangle-{\mathbb E}\langle R_{1,2} R_{2,3}\rangle-{\mathbb E}\langle R_{1,3} R_{2,3}\rangle-({\mathbb E}\langle R_{1,2}\rangle)^2]=0.
$$
Both are valid almost everywhere in the coupling constant space.
The replica symmetric Gibbs state gives  
$$
{\mathbb E}\langle R_{1,2} R_{2,3}\rangle={\mathbb E}\langle R_{1,3} R_{2,3}\rangle ={\mathbb E}\langle R_{1,2} R_{1,3}\rangle.
$$
Then the   following  relation between two kinds of variance 
$$2\lim_{L \rightarrow \infty} {\mathbb E}\langle {\Delta R_{1,2} }^2 \rangle=
3\lim_{L \rightarrow \infty} {\mathbb E}\langle {\delta R_{1,2} }^2 \rangle,
$$
is obtained. Lemma \ref{delta} implies
$$\lim_{L \rightarrow \infty} {\mathbb E}\langle {\Delta R_{1,2} }^2 \rangle=0.
$$
This completes the proof of Theorem \ref{MT}. $\Box$

\end{document}